\documentclass{appolb}
\usepackage{epsfig}

\def\bge{\begin{equation}}
\def\ene{\end{equation}}
\def\bg{\begin{eqnarray}}
\def\en{\end{eqnarray}}

\def\ubar{{\bar{u}}}
\def\dbar{{\bar{d}}}
\def\sbar{{\bar{s}}}

\def\bge{\begin{equation}}
\def\ene{\end{equation}}
\def\bg{\begin{eqnarray}}
\def\en{\end{eqnarray}}

\def\ubar{{\bar{u}}}
\def\dbar{{\bar{d}}}
\def\sbar{{\bar{s}}}

\begin{document}
\title{ASPECTS OF ANOMALOUS GLUE
\thanks{Presented at the 54th Cracow School of Theoretical
Physics, Zakopane, June 12-20 2014.}
}
\author{Steven D. Bass 
\address{Stefan Meyer Institute for Subatomic Physics, \\
Austrian Academy of Sciences,
Boltzmanngasse 3, 1090 Vienna, Austria}
}
\maketitle
\begin{abstract}
Non-perturbative glue associated with gluon topology 
plays a vital role in determining the mass of the $\eta'$ 
meson. We give an introduction to axial U(1) physics and
explain how this non-perturbative glue also contributes 
to $\eta'$ interactions with other hadrons. 
We concentrate on resonant $\eta' \pi$ production in partial waves with exotic quantum numbers not compatible with a 
quark-antiquark state and on the $\eta'$ in nuclear media 
where there are new results from the COMPASS and CBELSA/TAPS
experiments respectively.
\end{abstract}

\section{Introduction}

The physics of the $\eta'$ is associated with OZI violation.
While pions and kaons are would-be Goldstone bosons associated 
with chiral symmetry the isosinglet $\eta$ and $\eta'$ mesons 
are too massive by about 300-400 MeV for them to be pure 
Goldstone states.
They receive extra mass from non-perturbative gluon dynamics associated with the QCD axial anomaly;
for recent reviews see \cite{shore,cracow13}.
This non-perturbative glue is also expected to influence the 
$\eta'$-nucleon interaction \cite{bass99},
the behaviour of $\eta$ and $\eta'$ mesons in nuclear media \cite{etaA} as well as $\eta'$ production and decay 
processes [5-9].
In this paper we focus on the $\eta'$ and 
non-perturbative gluon dynamics in $\eta'$ phenomenology 
with emphasis on the $\eta'$ in the nuclear medium and 
resonant exclusive $\eta' \pi^-$ production where there 
are fresh data from experiments at ELSA in Bonn and 
COMPASS at CERN respectively.
New experiments on the $\eta'$ in nuclei are planned 
or underway at ELSA and GSI/FAIR.

Without the gluonic mass contribution the $\eta'$ and 
$\eta$ mesons would be strange and light-quark systems, 
like the flavour structure of the isoscalar $\phi$ and 
$\omega$ vector mesons.  
To the extent that interactions of the $\eta'$ 
with nucleons and nuclei are induced 
by light-quark components in the $\eta'$, 
then any observed $\eta'$-nucleon scattering length 
or $\eta'$ mass shift in nuclei is induced by the 
same non-perturbative glue that generates the large 
$\eta'$ mass.
Interesting new data from the CBELSA/TAPS Collaboration 
suggest an $\eta'$ mass shift of about -37 MeV at nuclear
matter density \cite{nanova},
very close to the prediction of the Quark Meson Coupling 
model \cite{etaA}. 
The first measurement of the $\eta'$-nucleon scattering
length in free space has recently been obtained 
by the COSY-11 experiment \cite{Czerwinski:2014yot}.
The $\eta'$ has also been suggested to play a prominent 
feature in constituent quark model calculations \cite{plessas} for stabilising the baryon spectra.

In $\eta' \pi$ rescattering processes one finds a strong 
OZI violating contribution to the odd $L$ partial wave contribution with exotic quantum numbers $L^{-+}$ 
\cite{bassmarco},
a resonant contribution that cannot be described by a 
simple quark-antiquark state. 
A factor of 5-10 enhancement is observed for $\eta' \pi^-$
production over $\eta \pi^-$ production in the $L=1,3,5$ 
partial waves in exclusive production at COMPASS 
\cite{compassexotic}.
Large branching ratios 
for $B$ and $D_s$ meson decays
to $\eta'$ final states observed at $B$-factories
are believed to be mediated by strong coupling to gluonic intermediate states \cite{frere,fritzsch}.

Further, the physics of the flavour-singlet QCD axial anomaly 
is important in understanding the spin structure of the proton
[14-18].
The nucleon's flavour-singlet axial-charge
which measures the quark spin content of the proton 
also picks up a polarised gluon contribution
$-3 {\alpha_s \over 2 \pi} \Delta g$ \cite{etar}
as well as a 
possible topological contribution 
with support only at Bjorken $x$ equal to zero associated
with a possible subtraction constant in the dispersion relation for the nucleon's $g_1$ spin dependent structure function measured in polarised deep inelastic scattering 
\cite{SBrmp1,bassmpla}.
The quest to measure this polarised glue has inspired vast
experimental activity at CERN, DESY and RHIC. 
The present status is that the gluon polarisation at 
the scale of the experiments, about 3-10 GeV$^2$, 
is finite \cite{vogelsang} and probably less than about 0.5,
consistent with the expectation from gluon radiation 
from valence quarks in the nucleon \cite{Bass:1998rn}.

The plan of this paper is as follows. 
In Section 2 we give a brief introduction to the $\eta'$ 
mass problem. 
Section 3 describes the QCD axial anomaly and ideas about 
non-perturbative anomalous glue (glue connected to the anomaly)
that might explain the $\eta'$ mass.
Then, in Section 4 we explain how this anomalous glue enters
in low-energy QCD processes involving 
the $\eta'$ in meson production and in the nuclear medium.
Finally, we summarise and 
conclude in Section 5 with an outlook to future developments.

\section{The $\eta'$ mass problem}

Low energy QCD is characterised by confinement and dynamical
chiral symmetry breaking.
The absence of parity doublets in the hadron spectrum tells 
us that the near-chiral symmetry for light $u$ and $d$ quarks 
is spontaneously broken. 
Scalar confinement implies dynamical chiral symmetry breaking. 
For example, in the Bag model the Bag wall connects left and right handed quarks leading to quark-pion coupling and the pion cloud of the nucleon \cite{awtcbm}.
Spontaneous chiral symmetry breaking in QCD is associated with a non-vanishing chiral condensate
\begin{equation}
\langle \ {\rm vac} \ | \ {\bar \psi} \psi \ | \ {\rm vac} \ \rangle < 0
.
\label{eq5}
\end{equation}
This spontaneous symmetry breaking induces an octet of 
Goldstone bosons associated with SU(3) and also 
(before extra gluonic effects in the singlet channel)
a flavour-singlet Goldstone boson.
The Goldstone bosons $P$ couple to the axial-vector currents 
which play the role of Noether currents through
\begin{equation}
\langle {\rm vac} | J_{\mu 5}^i | P(p) \rangle = 
-i f_P^i \ p_{\mu} e^{-ip.x}
\end{equation}
with $f_P^i$ the corresponding decay constants
and satisfy the Gell-Mann-Oakes-Renner relation
\begin{equation}
m_{\pi}^2 f_{\pi}^2 = - m_q \langle {\bar \psi} \psi \rangle .
\end{equation}
The pion and kaon fit well in this picture.

The isosinglet $\eta$ and $\eta'$ masses are about 300-400 MeV 
too heavy to be pure Goldstone states.
One needs extra mass in the flavour-singlet channel associated 
with non-perturbative topological gluon configurations 
\cite{shore,crewther}, related perhaps to confinement [23-27] or instantons \cite{thooft}.
SU(3) breaking generates mixing between the octet and 
singlet states yielding the massive $\eta$ and $\eta'$ bosons.

To see the effect of the gluonic mass contribution consider the $\eta$-$\eta'$ mass matrix for free mesons
(at leading order in the chiral expansion) 
\begin{equation}
M^2 =
\left(\begin{array}{cc}
{4 \over 3} m_{\rm K}^2 - {1 \over 3} m_{\pi}^2  &
- {2 \over 3} \sqrt{2} (m_{\rm K}^2 - m_{\pi}^2) \\
\\
- {2 \over 3} \sqrt{2} (m_{\rm K}^2 - m_{\pi}^2) &
[ {2 \over 3} m_{\rm K}^2 + {1 \over 3} m_{\pi}^2 + {\tilde m}^2_{\eta_0} ]
\end{array}\right)
.
\label{eq10}
\end{equation}
Here ${\tilde m}^2_{\eta_0}$ 
is the flavour-singlet gluonic mass term.

The masses of the physical $\eta$ and $\eta'$ mesons are found
by diagonalising this matrix, {\it viz.}
\begin{eqnarray}
| \eta \rangle &=&
\cos \theta \ | \eta_8 \rangle - \sin \theta \ | \eta_0 \rangle
\\ \nonumber
| \eta' \rangle &=&
\sin \theta \ | \eta_8 \rangle + \cos \theta \ | \eta_0 \rangle
\label{eq11}
\end{eqnarray}
where
\begin{equation}
\eta_0 = \frac{1}{\sqrt{3}}\; (u\ubar + d\dbar + s\sbar),\quad
\eta_8 = \frac{1}{\sqrt{6}}\; (u\ubar + d\dbar - 2 s\sbar) 
.
\label{mixing2}
\end{equation}
One obtains values for the $\eta$ and $\eta'$ masses
\begin{equation}
m^2_{\eta', \eta} 
= (m_{\rm K}^2 + {\tilde m}_{\eta_0}^2 /2)
\pm {1 \over 2}
\sqrt{(2 m_{\rm K}^2 - 2 m_{\pi}^2 - {1 \over 3} 
{\tilde m}_{\eta_0}^2)^2 + {8 \over 9} {\tilde m}_{\eta_0}^4} 
.
\label{eq12}
\end{equation}

The gluonic mass term is obtained from summing over 
the two eigenvalues in Eq.(7) 
to give the Witten-Veneziano mass formula \cite{witten,vecca}
\begin{equation}
m_{\eta}^2 + m_{\eta'}^2 = 2 m_K^2 + {\tilde m}_{\eta_0}^2 .
\end{equation}
Substituting the physical values of 
$m_{\eta}$, $m_{\eta'}$ and  $m_K$ 
gives ${\tilde m}_{\eta_0}^2 = 0.73$GeV$^2$.
The gluonic mass term has a rigorous interpretation 
in terms of the Yang-Mills topological susceptibility, 
see Eqs.(10-12) below.

In the OZI limit of no gluonic mass term
the $\eta$ would be approximately an isosinglet light-quark state
(${1 \over \sqrt{2}} | {\bar u} u + {\bar d} d \rangle$)
with mass $m_{\eta} \sim m_{\pi}$
degenerate with the pion and
the $\eta'$ would be a strange-quark state $| {\bar s} s \rangle$
with mass $m_{\eta'} \sim \sqrt{2 m_{\rm K}^2 - m_{\pi}^2}$
--- mirroring the isoscalar vector $\omega$ and $\phi$ mesons.

Phenomenological studies of various decay processes give a value for the $\eta$-$\eta'$ mixing angle between 
$-15^\circ$ and $-20^\circ$ 
[6,31-33].
%
This mixing means that non-perturbative glue through axial 
U(1) dynamics plays an important role in both the $\eta$ and $\eta'$ and their interactions.
Treating the $\eta$ as an octet 
pure would-be Goldstone boson risks losing essential physics.
Recent lattice calculations give values for the mixing angle 
between about $-10^\circ$ and $-20^\circ$ \cite{latticetheta}.

\section{Non-perturbative glue and ${\tilde m}_{\eta_0}^2$}

The gluonic mass term is related to the QCD axial anomaly 
in the divergence of the flavour-singlet axial-vector current
\cite{adler}.
While the non-singlet axial-vector currents are partially conserved (they have just mass terms in the divergence), 
the singlet current
$
J_{\mu 5} = \bar{u}\gamma_\mu\gamma_5u
+ \bar{d}\gamma_\mu\gamma_5d + \bar{s}\gamma_\mu\gamma_5s 
$
satisfies the anomalous divergence equation 
\begin{equation}
\partial^\mu J_{\mu5} = 6 Q
+ \sum_{k=1}^{3} 2 i m_k \bar{q}_k \gamma_5 q_k 
\end{equation}
where 
$
Q = {\alpha_s \over 8 \pi} G_{\mu \nu} {\tilde G}^{\mu \nu}
$
is the topological charge density,
$G_{\mu \nu}$ is the gluon field tensor and
${\tilde G}^{\mu \nu} = 
{1 \over 2} \epsilon^{\mu \nu \alpha \beta} G_{\alpha \beta}$.
The axial anomaly derives from a clash of symmetries.
In classical field theory 
the axial-vector currents are always partially conserved.
In quantum field theory the flavour-singlet axial-vector
current can couple through gluon intermediate states.
Here the axial-vector current, two vector quark-gluon-vertex
triangle diagram is essential. 
When we regularise the ultraviolet behaviour of momenta in
the loop, we find that we can preserve current conservation
at the quark-gluon-vertices (necessary for gauge invariance and renormalisability) or partial conservation of the axial-vector current but not both simultaneously.
Current conservation wins and induces 
the gluonic anomaly term in the singlet divergence equation
(9) from the pointlike ultraviolet part of the triangle loop.

Besides adding the extra term to the RHS of the divergence equation (9), the anomaly opens a window to non-perturbative gluon dynamics.
The integral over space 
$\int \ d^4 z \ Q = n$ 
measures the gluonic winding number \cite{crewther} 
as a measure of non-local topological structure
which is an integer for (anti-)instantons and which vanishes in perturbative QCD.
The QCD topological charge plays a vital role in the $\eta'$.

The Witten-Veneziano mass formula \cite{witten,vecca} 
for the $\eta'$ mass
follows from the flavour-singlet Ward identities and
relates the gluonic mass term for the singlet boson to the topological susceptibility of pure Yang-Mills (glue with no quarks),
{\it viz.}
\begin{equation}
{\tilde m}_{\eta_0}^2 
= - {6 \over f_{\pi}^2} \chi (0)|_{\rm YM} ,
\end{equation}
where
\begin{equation}
\chi (k^2)|_{\rm YM} = 
\int d^4 z \ i \ e^{ik.z} \
\langle {\rm vac} | \ T \ Q(z) Q(0) \ | {\rm vac} \rangle 
\big|_{\rm YM} .
\end{equation}
Understanding the QCD dynamical origin of this topological
susceptibility (which gluon configurations saturate it) 
and how it contributes to hadron phenomenology are at heart of axial U(1) physics.
In the QCD limit of large number of colours, $N_c$, 
if we assume that the topological winding number 
remains finite independent of the value of $N_c$
then
\begin{equation}
{\tilde m}_{\eta_0}^2 \sim 1 / N_c .
\end{equation}
The topological susceptibility $\chi(0)$ in full QCD 
(with quarks) vanishes in the limit of massless quarks
where the pure Yang-Mills contribution is cancelled against
a pole term involving the massive $\eta'$.
QCD lattice calculations 
give values 
$\chi^{1/4} (0)|_{\rm YM} = 191 \pm 5$ MeV \cite{giusti}
and
$\chi^{1/4} (0)|_{\rm YM} = 193 (1)(8)$ MeV \cite{hoelbling},
very close to the value 180 MeV 
which follows from taking
${\tilde m}_{\eta_0}^2 = 0.73$GeV$^2$
in the Witten-Veneziano formula Eq.(8).

The topological charge density $Q$ is a total divergence.
The anomaly equation can also be written
\begin{equation}
\partial^\mu J_{\mu5} = 
6 \partial^\mu K_\mu 
+
\sum_{k=1}^{3} 2 i m_k \bar{q}_k \gamma_5 q_k 
\label{eqf102}
\end{equation}
where
\begin{equation}
K_{\mu} = {g^2 \over 32 \pi^2}
\epsilon_{\mu \nu \rho \sigma}
\biggl[ A^{\nu}_a \biggl( \partial^{\rho} A^{\sigma}_a
- {1 \over 3} g
f_{abc} A^{\rho}_b A^{\sigma}_c \biggr) \biggr]
\label{eqf103}
\end{equation}
is the 
gluonic Chern-Simons current and $\alpha_s = g^2/4 \pi$
is the QCD coupling.

Eq.(\ref{eqf102}) allows us to define a partially conserved current
\begin{equation}
J_{\mu 5} = J_{\mu 5}^{\rm con} + 2f K_{\mu} \ 
\end{equation}
with
\begin{equation}
\partial^\mu J^{\rm con}_{\mu5}
= \sum_{i=1}^{3} 2im_i \bar{q}_i\gamma_5 q_i \ .
\end{equation}
When we make a gauge transformation $U$ the gluon field transforms as
\begin{equation}
A_{\mu} 
\rightarrow U A_{\mu} U^{-1} + {i \over g} (\partial_{\mu} U) U^{-1}
\end{equation}
and the operator $K_{\mu}$ transforms as
\begin{eqnarray}
K_{\mu} \rightarrow K_{\mu}
&+&
 i {g \over 8 \pi^2} \epsilon_{\mu \nu \alpha \beta}
\partial^{\nu}
\biggl( U^{\dagger} \partial^{\alpha} U A^{\beta} \biggr)
\nonumber \\
&+& 
{1 \over 24 \pi^2} \epsilon_{\mu \nu \alpha \beta}
\biggl[
(U^{\dagger} \partial^{\nu} U)
(U^{\dagger} \partial^{\alpha} U)
(U^{\dagger} \partial^{\beta} U)
\biggr]
\label{eqf106}
\end{eqnarray}
where the third term on the RHS is associated 
with gluon gauge field topology \cite{crewther}.
Before possible confinement considerations \cite{witteninst},
finiteness of the QCD action requires that 
$ x^2 G_{\mu \nu} \rightarrow 0 $
as $x_{\mu} \rightarrow \infty$ in almost every direction
so the field $A_{\mu}$ should tend to a pure gauge configuration
$ g A_{\mu} \rightarrow i G^{-1} \partial_{\mu} G $.
With this constraint
\begin{equation}
\nu = \int d^4 x \ Q
= \int d^4 x \ \partial^{\mu} K_{\mu}
= \int \sigma^{\mu} K_{\mu}
= {\rm integer}
\end{equation}
for instanton topological charge \cite{crewther}. 
A finite topological charge comes here from the requirement that the gluon field is pure gauge at infinity and that the gauge group is topologically non-trivial.
The topological winding number is determined by the gluonic 
boundary conditions at ``infinity'' (a large surface with boundary which is spacelike with respect to the positions 
$z_k$ of any operators or fields in the physical problem). 
It is insensitive to local deformations of the gluon
field $A_{\mu}(z)$ or of the gauge transformation $U(z)$.

The current $J_{\mu 5}$ in QCD is multiplicatively renormalised and picks up a two loop anomalous dimension from the axial anomaly \cite{crewther,kodaira}.
(Partially) conserved currents are not renormalised.
It follows that $J_{\mu 5}^{\rm con}$ is renormalisation scale invariant and the scale dependence of $J_{\mu 5}$ 
is carried entirely by $K_{\mu}$.
Gauge transformations shuffle a scale invariant operator 
quantity between the two operators $J_{\mu 5}^{\rm con}$ 
and $K_{\mu}$ whilst keeping $J_{\mu 5}$ invariant.

In general, matrix elements of $K_{\mu}$ are gauge dependent.
This means that one has to be careful writing matrix elements 
of $J_{\mu 5}$ as the sum of
(measurable) ``quark'' and ``gluonic'' contributions.
For the $\eta'$, if we consider the singlet boson version of 
Eq.(2) and write $J_{\mu 5}$ as the sum of 
$J_{\mu 5}^{\rm con}$ and $K_{\mu}$ contributions,
then the two matrix elements corresponding to the 
gauge dependent currents 
are in general separately each gauge dependent.
Isolating a gluonic leading Fock component from the $\eta'$ 
involves subtle issues of gauge invariance and only makes
sense with respect to a particular renormalisation scheme 
like the gauge invariant scheme 
$\overline{\rm MS}$ \cite{sbcracow}.

Since the topological charge density is a total divergence it
follows that the gluonic Chern-Simons current (14) should 
couple to a massless pole. 
This is the Kogut-Susskind pole \cite{ks}.
The fact that there is no massless flavour-singlet Goldstone 
boson in the physical spectrum implies that this pole 
cancels against a second massless pole term which couples 
to the partially conserved current $J_{\mu 5}^{\rm con}$ 
with equal residue and opposite sign. 
This second pole term stays massless when quark masses 
are turned on to cancel the pole term coupled to $K_{\mu}$.

Corresponding to the two currents 
$J_{\mu 5}$ and $J_{\mu 5}^{\rm con}$, 
we can define the two operator charges
$ X(t) = \int d^3z \ J_{0 5} (z) $
and
$ Q_5  = \int d^3z \ J_{0 5}^{\rm con}(z)$.
The charge $X(t)$ is manifestly gauge invariant whereas 
$Q_5$ is invariant only under ``small'' gauge transformations
which are topologically deformable to the identity.
Choosing the gauge $A_0=0$, 
the charge $Q_5$ transforms 
as $ Q_5 \rightarrow Q_5 - 2 f n $
where $n$ is the winding number associated with the gauge transformation $U$. 
Although $Q_5$ is gauge dependent we can define a gauge invariant chirality $q_5$ for a given operator ${\cal O}$ through the gauge-invariant eigenvalues of the equal-time commutator
$
[ \ Q_5 \ , \ {\cal O} \ ]_{-} = - q_5 \ {\cal O} 
$.
The gauge invariance of $q_5$ follows because 
this commutator appears in gauge invariant Ward Identities 
despite the gauge dependence of $Q_5$ \cite{crewther}. 
One finds that the time derivative of spatial components of 
the gluon field have zero chirality $q_5$
(following from the non-renormalisation of 
 the partially conserved current $J_{\mu 5}^{\rm con}$) 
but non-zero $X$ charge, which is induced by the anomaly.
The analogous situation in QED 
is discussed in Refs. \cite{adler,boulware,brandeis}.
If one requires that chirality is renormalisation group invariant and that the time derivative of 
the spatial components of the gluon field have zero chirality, 
then one is led to using $q_5$ associated 
with $J_{\mu 5}^{\rm con}$ to define chirality.

When topological effects are taken into account, the QCD vacuum $|\theta \rangle$ is understood to be a coherent Bloch superposition of states characterised by different topological winding number $m$
\cite{callan} 
\begin{equation}
| {\rm vac}, \theta \rangle = \sum_m e^{i m \theta}
 \  | m \rangle 
\label{eqf119}
\end{equation}
where the QCD $\theta$ angle is experimentally less than 
$10^{-10}$ \cite{pdg}.
For integer values of the topological winding number $m$,
the states $|m \rangle$ contain $mf$ quark-antiquark 
pairs with non-zero $Q_5$ chirality
$\sum_l \chi_l = - 2 f m$ 
where $f$ is the number of light-quark flavours.
Relative to the $|m=0 \rangle$ state, the $|m=+1 \rangle$ 
state carries topological winding number +1 and $f$
quark-antiquark pairs with $Q_5$ chirality equal to $-2f$.
Each state $|m \rangle$ carries zero net axial-charge as measured by $J_{\mu 5}$ and $X(t)$.

Invariance under ``large'' gauge transformations 
which change the topological winding number is like saying 
the physics is invariant under the choice of which state 
$|m \rangle$ we choose as ``zero'' when we set up a 
``ruler'' to label the $\theta$ vacuum states $|m \rangle$.
A ``large'' gauge transformation which changes the topological
charge $m$ by $k$ units just changes the phase of the 
$\theta$-vacuum by amount $e^{-ik \theta}$.

Instanton tunneling processes between different $|m \rangle$ states differing by topological winding number one,
$|m \rangle \rightarrow |m \pm 1 \rangle$,
connect left and right handed fermions \cite{thooft}. 
A flavour-singlet combination of quarks incident on an instanton ``vertex'' annihilate (fill up vacant levels) 
with the delocalised quark-antiquark pairs 
in the neighbouring state, 
thus liberating a flavour-singlet combination of quarks with 
opposite chirality into the final state. 
Energy-momentum is conserved between incoming and outgoing quarks with net axial charge conserved 
by $q_5$ chirality being absorbed into the ``vacuum'', 
that is shifted into a zero-mode.
This process can induce a topological $x=0$ contribution 
to the proton's flavour-singlet axial-charge or 
``quark spin content'' \cite{bassmpla}
depending on whether instantons
spontaneously \cite{crewther} or explicitly \cite{thooftrep} break axial U(1) symmetry. 
The valence quarks can thus act as a source for polarising the $\theta$ vacuum inside a proton relative to the vacuum 
outside by shifting some fraction of the spin of moving partons into a topological zero-mode contribution 
\cite{SBrmp1,bassmpla}.
The proton matrix element of the abelian part 
(first term) of the gluonic Chern-Simons current
in light-cone co-ordinate, $K_+$, in $A_+=0$ gauge
measures the gluon polarisation in the nucleon \cite{etar}.

Instantons are an important source of 
finite topological charge and come with large $N_c$ 
behaviour $\sim e^{-N_c}$, 
decaying faster with increasing $N_c$
than ${\tilde m}_{\eta_0}^2 \sim 1/N_c$
in Eq.(12).
Witten argued that with quark confinement the vacuum 
no longer requires gluon fields to be pure gauge at 
infinity.
One might also get non-vanishing topological charge 
from non-instanton effects.
Additional gluonic configurations with topological 
charge suggested 
in the literature 
involve calorons \cite{calerons}
and centre-vortices \cite{topsusc,centrev}.
Large $N_c$ and more quark model like approaches 
to understanding the $\eta'$ wavefunction
involving non-perturbative gluon intermediate states 
\cite{mink}
might here be reconciled if
{\it e.g.} confinement processes associated with topology
induce a dynamical scalar component in 
the non-perturbative quark gluon vertex connecting 
left- and right- handed quarks in the flavour-singlet 
channel in addition to effects associated with pion and
kaon production in the non-singlet channels \cite{alkofer}.

This formalism generalises readily to the definition of 
baryon number in the presence of electroweak gauge fields 
\cite{Bass:2004}.
The vector current which measures baryon number is sensitive 
to the axial anomaly through the parity violating electroweak interactions.
This vector current can be written as the sum of left and right
handed currents:
\begin{equation}
J_{\mu}
= {\bar \Psi} \gamma_{\mu} \Psi
= {\bar \Psi} \gamma_{\mu} {1 \over 2} (1 - \gamma_5) \Psi
+ {\bar \Psi} \gamma_{\mu} {1 \over 2} (1 + \gamma_5) \Psi
.
\label{eq3}
\end{equation}
In the Standard Model the left handed fermions couple 
to the SU(2) electroweak gauge fields $W^{\pm}$ and $Z^0$.
This means that this baryon current is sensitive to the axial anomaly \cite{thooft}.
One finds the anomalous divergence equation
\begin{equation}
\partial^\mu J_{\mu}
= n_f
\biggl( - \partial^\mu K_\mu + \partial^\mu k_\mu
\biggr)
\label{eq4}
\end{equation}
where $K_{\mu}$ and $k_{\mu}$ are the SU(2) electroweak 
and U(1) hypercharge anomalous Chern-Simons currents.
If one requires that baryon number is renormalisation group invariant and that the time derivative of the spatial components of the W boson field have zero baryon number, 
then one is led to using the conserved vector current 
analogy of $q_5$ to define the baryon number \cite{Bass:2004}, 
{\it e.g.} in Sphaleron induced electroweak baryogenesis in 
the early Universe \cite{sphaleron}.

\section{The $\eta'$ in low-energy QCD}

Independent of the detailed QCD dynamics one can construct 
low-energy effective chiral Lagrangians which include the 
effect of the anomaly and axial U(1) symmetry, 
and use these Lagrangians to study low-energy processes involving the $\eta$ and $\eta'$.

The physics of axial U(1) degrees of freedom is described 
by the U(1)-extended low-energy effective Lagrangian 
\cite{veccc}.
In its simplest form this reads
\begin{eqnarray}
{\cal L} =
{F_{\pi}^2 \over 4}
{\rm Tr}(\partial^{\mu}U \partial_{\mu}U^{\dagger})
+
{F_{\pi}^2 \over 4} {\rm Tr} M \biggl( U + U^{\dagger} \biggr)
\nonumber \\
+ {1 \over 2} i Q {\rm Tr} \biggl[ \log U - \log U^{\dagger} \biggr]
+ {3 \over {\tilde m}_{\eta_0}^2 F_{0}^2} Q^2
.
\label{eq20}
\end{eqnarray}
Here 
$
U = \exp \ i \biggl(  \phi / F_{\pi}
                  + \sqrt{2 \over 3} \eta_0 / F_0 \biggr) 
$
is the unitary meson matrix where 
$\phi = \ \sum \pi_a \lambda_a$ 
denotes the octet of would-be Goldstone bosons associated 
with spontaneous chiral $SU(3)_L \otimes SU(3)_R$ breaking
and $\eta_0$ is the singlet boson.
In Eq.(23) $Q$ denotes the topological charge density;
$M = {\rm diag} [ m_{\pi}^2, m_{\pi}^2, 2 m_K^2 - m_{\pi}^2 ]$
is the quark-mass induced meson mass matrix.
The pion decay constant $F_{\pi} = 92.4$MeV and 
$F_0$ is the flavour-singlet decay constant,
$F_0 \sim F_{\pi} \sim 100$ MeV \cite{gilman}.

The flavour-singlet potential involving $Q$ is introduced to generate the gluonic contribution to the $\eta$ and $\eta'$ masses and to reproduce the anomaly in the divergence of
the gauge-invariantly renormalised flavour-singlet 
axial-vector current.
The gluonic term $Q$ is treated as a background field with no kinetic term. It may be eliminated through its equation of motion to generate a gluonic mass term for the singlet boson,
{\it viz.}
\begin{equation}
{1 \over 2} i Q {\rm Tr} \biggl[ \log U - \log U^{\dagger} \biggr]
+ {3 \over {\tilde m}_{\eta_0}^2 F_{0}^2} Q^2
\
\mapsto \
- {1 \over 2} {\tilde m}_{\eta_0}^2 \eta_0^2
.
\label{eq23}
\end{equation}
The most general low-energy effective Lagrangian involves a 
$U_A(1)$ invariant polynomial in $Q^2$. Higher-order terms in $Q^2$ become important when we consider scattering processes involving more than one $\eta'$ \cite{veccb}.
In general, couplings involving $Q$ give OZI violation in physical observables.
The interactions of the $\eta$ and $\eta'$ with other mesons and with nucleons can be studied by coupling the Lagrangian Eq.(23) to other particles.

\subsection{Light-mass exotic meson production}

The OZI violating interaction
\begin{equation}
{\cal L}_{m2Q} =
\lambda Q^2 \partial_{\mu} \pi_a \partial^{\mu} \pi_a
\end{equation}
is needed to generate the leading (tree-level)
contribution to the decay $\eta' \rightarrow \eta \pi \pi$
\cite{veccb}.
When iterated in the Bethe-Salpeter equation for meson-meson
rescattering
this interaction yields a dynamically generated 
exotic state with quantum numbers $J^{PC} = 1^{-+}$ 
and mass about 1400 MeV 
in $\eta' \pi$ rescattering mediated by the OZI violating coupling of the $\eta'$ \cite{bassmarco}.

This suggests a dynamical interpretation of the light-mass 
$1^{-+}$ exotics observed in experiments at BNL \cite{exoticb} and CERN \cite{exoticc}.
These mesons are particularly interesting because the quantum numbers $J^{PC}=1^{-+}$ are inconsistent with a simple 
quark-antiquark bound state.

Further, this OZI violating interaction will also 
play an important role in higher $L$ odd partial waves.
Different partial wave contributions are projected out 
via integrating
\begin{equation}
T_L =  \frac{1}{2} 
\int_{-1}^{+1} d (\cos \theta)  \ P_L (\cos \theta) \ T
\end{equation}
weighted by Legendre polynomials $P_L$ with $T$ the T matrix.
The Legendre polynomials are either odd or even in 
$\cos \theta$ about zero. 
So if we get an effect in $L=1$ (p-wave), 
it must come from something odd in $\cos \theta$ in $T$. 
This odd contribution will also integrate to finite values with $L=3, 5, ...$ (not $L=2,4,...$) 
with the key determining feature being whether the driving 
term is odd or even about zero in $\cos \theta$.

The COMPASS experiment at CERN has just measured exclusive production of $\eta' \pi^-$ and $\eta \pi^-$ 
in 191 GeV $\pi^-$ collisions from a Hydrogen target
\cite{compassexotic}.
They find the interesting result that
$\eta' \pi^-$ production 
is enhanced relative to $\eta \pi^-$ production 
by a factor of 5-10 in the exotic $L=1,3,5$ partial waves 
with quantum numbers $L^{-+}$ in the inspected invariant 
mass range up to 3 GeV. No enhancement was observed in the
even $L$ partial waves.

\subsection{The $\eta'$ in nuclear matter}

Measurements of the $\eta-$ and $\eta'-$ (as well as pion and 
kaon) nucleon and nucleus systems promise to yield valuable new information about dynamical chiral and axial U(1) symmetry breaking in low energy QCD. The quark condensate is modified in the nuclear environment which leads to changes in the properties of hadrons in medium including the masses of the Goldstone bosons \cite{kienle}.
How does the gluonic mass contribution to the $\eta$ and $\eta'$ change in nuclei ?

With increasing density chiral symmetry is partially restored corresponding to a reduction in the value of the quark condensate and pion decay constant $f_{\pi}$.
Experiments with pionic atoms give a value
$f_{\pi}^{*2}/f_{\pi}^2 = 0.64 \pm 0.06$ 
at nuclear matter density $\rho_0$ \cite{kienle,suzuki}.
This implies changes in the meson masses in the medium and 
also the coupling of the Goldstone bosons to the constituent 
quarks and the nucleon. 
For reviews of medium modifications at finite density and
the QCD phase diagram see [55-57].
These medium modifications need to be understood 
self-consistently within the interplay of confinement,
spontaneous chiral symmetry breaking and axial U(1) dynamics.
In the limit of chiral restoration the pion should decouple 
from the physics, the pion decay constant $f_{\pi}$ go to zero and (perhaps) with scalar confinement the pion 
constituent-quark and pion nucleon coupling constants should vanish with dissolution of the pion wavefunction.

For pions one finds a small mass shift of order a few MeV in nuclear matter \cite{kienle}
whereas kaons are observed to experience an effective 
mass drop for the $K^-$ to about 270 MeV at two times 
nuclear matter density in heavy-ion collisions \cite{kaons}.

The $\eta$ and $\eta'$-nucleon interactions are believed to 
be attractive suggesting that these mesons may form 
strong-interaction bound-states in nuclei.
For the $\eta$ one finds a sharp rise at threshold 
in the cross section 
for $\eta$ production from Helium in 
photoproduction \cite{mamieta}
and proton-deuteron collisions \cite{cosyeta}
which may hint at a reduced $\eta$ mass in the nuclear medium.
The $\eta'$ is very interesting. 
As we have seen in Section 2, 
without glue this would be a strange quark state. 
To the extent that coupling to 
nucleons and nuclear matter (e.g. via the $\sigma$ 
correlated two-pion mean field in the nucleus)
is induced by light-quark components in the meson,
any observed $\eta'$-nucleon scattering length and 
$\eta'$ mass shift in medium is induced by the QCD 
axial anomaly that generates 
part of the $\eta'$ mass \cite{cracow13}.

Meson mass shifts can be investigated through bound 
state searches in nuclei. 
Meson mass shifts can also be investigated through studies of excitation functions in photoproduction experiments from 
nuclear targets. 
The production cross section is enhanced with the lower effective mass in the nuclear medium. 
When the meson leaves the nucleus it returns on-shell 
to its free mass with the energy budget conserved at
the expense of the kinetic energy so that excitation functions
and momentum distributions can provide essential clues to the
meson properties in medium \cite{metag}.
Using this physics a first (indirect) estimate of the $\eta'$ 
mass shift has recently been deduced 
by the CBELSA/TAPS Collaboration \cite{nanova}.
The $\eta'$-nucleus optical potential 
$V_{\rm opt} = V_{\rm real} + iW$
deduced from these photoproduction experiments is 
\begin{eqnarray}
V_{\rm real} (\rho_0)
= m^* - m 
&=& -37 \pm 10 (stat.) \pm 10 (syst.) \ {\rm MeV}
\nonumber \\ 
W(\rho_0) &=& -10 \pm 2.5 \ {\rm MeV}
\end{eqnarray}
at nuclear matter density.
These numbers with small imaginary part \cite{elsa}
suggest that possible $\eta'$ bound states in nuclei
may be within reach of forthcoming experiments. 
For clean observation of a bound state one needs the real part of the optical potential to be much bigger than the imaginary part.
The mass shift, Eq.(27), is also very similar to the expectations of the Quark Meson Coupling model, see below.

There is presently a vigorous experimental programme to search 
for evidence of $\eta$ and $\eta'$ bound states with ongoing experiments at COSY to look for possible $\eta$ bound states in Helium 
\cite{pawela}
and new experiments in photoproduction at ELSA \cite{elsap}
and using the (p, d) reaction at GSI/FAIR \cite{gsi}
to look for possible $\eta'$ bound states in Carbon.
Eta bound states in Helium require a large $\eta-$nucleon 
scattering length with real part greater than about 0.9 fm 
\cite{gal}.

Medium modifications can be understood at the quark level 
through coupling of the scalar isoscalar $\sigma$ 
(and also $\omega$ and $\rho$) mean fields in the nucleus 
to the light quarks in the hadron \cite{qmc}.
The binding energies and in-medium masses of the $\eta$ and $\eta'$ are sensitive to the flavour-singlet component in 
the mesons and hence to the non-perturbative glue associated with axial U(1) dynamics \cite{etaA}.

Within the effective Lagrangian approach of Eq.(23)
the medium dependence of 
${\tilde m}_{\eta_0}^2$ is introduced through coupling 
to the $\sigma$ mean-field in the nucleus through the interaction term
\begin{equation}
{\cal L}_{\sigma Q} = g_{\sigma Q} \ Q^2 \ \sigma 
\end{equation}
where $g_{\sigma Q}$ denotes coupling to the $\sigma$ mean field. One finds the gluonic mass term decreases in-medium
${\tilde m}_{\eta_0}^{*2} < {\tilde m}_{\eta_0}^2$ 
independent of the sign of $g_{\sigma Q}$ and the medium acts
to partially neutralise axial U(1) symmetry breaking by gluonic effects \cite{etaA}.
To estimate the size of the effect we look to phenomenology and QCD motivated models.

Interesting results for the $\eta$ and $\eta'$ mass shifts 
with $\eta$-$\eta'$ mixing are obtained 
within the Quark Meson Coupling model (QMC)
of hadron properties in the nuclear medium 
\cite{etaA}.
Here the large $\eta$ and $\eta'$ masses are used to motivate 
taking an MIT Bag description for the meson wavefunctions.
Gluonic topological effects are understood to be ``frozen in'', meaning that they are only present implicitly through 
the masses and mixing angle in the model.
The in-medium mass modification comes from coupling the light 
(up and down) quarks and antiquarks in the meson wavefunction 
to the scalar $\sigma$ mean-field in the nucleus 
working in mean-field approximation \cite{qmc,qmcorig,etaqmc}.
The coupling constants in the model for the coupling of 
light-quarks to the $\sigma$ (and $\omega$ and $\rho$) 
mean-fields in the nucleus are adjusted to fit the saturation energy and density of symmetric nuclear matter and the bulk symmetry energy.
The strange-quark component of the wavefunction does not couple to the $\sigma$ field and $\eta$-$\eta'$ mixing is readily built into the model.
Gluon fluctuation and centre-of-mass effects are assumed to be independent of density.
The model results for the meson masses in medium and the real 
part of the meson-nucleon scattering lengths are shown in 
Table 1 for different values of the $\eta$-$\eta'$ mixing angle, 
which is taken 
to be density independent in these calculations \cite{etaA}.
The effective scattering length is calculated 
through
\begin{equation}
2 m \Delta m = 4 \pi \rho \ a \ (1 + m/M)
\end{equation}
where $m$ is the mass of the meson in free space,
$\Delta m$ is the mass shift in medium,
$\rho$ is the density of the nuclear medium, 
$M$ is the nucleon mass and $a$ is the scattering length
\cite{ericson}.
Increasing the flavour-singlet component in the $\eta$ at 
the expense of the octet component gives more attraction, 
more binding and a larger value of the $\eta$-nucleon scattering length, $a_{\eta N}$. 
The QMC model makes no claim about the imaginary part of the scattering length.

\begin{table}[t!]
\begin{center}
\caption{
Physical masses fitted in free space, the bag masses in medium at normal nuclear-matter density, $\rho_0 = 0.15$ fm$^{-3}$, 
and corresponding meson-nucleon scattering lengths.
}
\label{bagparam}
\begin{tabular}[t]{c|lll}
\hline
&$m$ (MeV) 
& $m^*$ (MeV) & ${\tt Re} a$ (fm)
\\
\hline
$\eta_8$  &547.75  
& 500.0 &  0.43 \\
$\eta$ (-10$^o$)& 547.75  
& 474.7 & 0.64 \\
$\eta$ (-20$^o$)& 547.75  
& 449.3 & 0.85 \\
$\eta_0$  &      958 
& 878.6  & 0.99 \\
$\eta'$ (-10$^o$)&958 
& 899.2 & 0.74 \\
$\eta'$ (-20$^o$)&958 
& 921.3 & 0.47 \\
\hline
\end{tabular}
\end{center}
\end{table}

For -20 degrees $\eta$-$\eta'$ mixing angle, QMC predicts 
the $\eta'$ mass shift to be -37 MeV at nuclear matter 
density $\rho_0$, 
corresponding to the real part of the effective 
$\eta'$-nucleon scattering length being 0.5 fm. 
This value is very similar to the mass shift 
$-37 \pm 10 \pm 10$ MeV deduced from photoproduction data 
by the CBELSA/TAPS Collaboration \cite{nanova}.
Further, COSY-11 have recently determined 
the $\eta'$-nucleon scattering length in free space 
to be
\begin{eqnarray}
\nonumber
\mathrm{Re}(a_{p\eta'}) &=&  0~\pm~0.43~\mathrm{fm}
\\
\mathrm{Im}(a_{p\eta'}) &=& 0.37^{~+0.40}_{~-0.16}~\mathrm{fm}\end{eqnarray}
from studies of the final state interaction 
in $\eta'$ production in proton-proton collisions 
close to threshold \cite{Czerwinski:2014yot}.
Larger mass shifts, 
downwards by up to 80-150 MeV, 
were found in Nambu-Jona-Lasinio model
\cite{hirenzaki}
and linear sigma model calculations \cite{jido1},
which in general also give a rising $\eta$ mass at 
finite density. 
Each of these theoretical models prefers a positive sign 
for the real part of $a_{\eta' N}$ in medium. 
The energy and density dependence of the $\eta'$- 
(and also $\eta$-) nucleon scattering lengths is a open
topic of investigation \cite{gal}.
If one assumes no density and energy dependence of the $\eta'$
nucleon scattering length, then the value obtained in Eq.(30) 
is consistent with the QMC result \cite{etaA} and 
disfavours the expectations in~\cite{hirenzaki,jido1}.
A chiral coupled channels calculation performed with
possible scattering lengths with real part 
between 0 and 1.5 fm is reported in \cite{osetetaprime}.

$\eta$-$\eta'$ mixing with the phenomenological mixing angle 
$-20^\circ$ leads to a factor of two increase 
in the mass-shift and 
in the scattering length obtained in the model
relative to the prediction for a pure octet $\eta_8$.
This result may explain why values of $a_{\eta N}$ 
extracted from phenomenological fits to experimental 
data where the $\eta$-$\eta'$ mixing angle is unconstrained \cite{wycech} give larger values than those predicted 
in theoretical coupled channels models where the $\eta$ 
is treated as a pure octet state \cite{etaweise,etaoset}. 
We refer to Ref. \cite{cracow13} for comparison of different
models and their predictions for the $\eta$ and $\eta'$
nucleus systems.

\section{Summary and Conclusions}

Besides generating the large $\eta'$ mass, 
non-perturbative 
anomalous glue plays an important role in driving $\eta'$
interactions and production and decay processes with significant OZI violation.
Glue associated with the QCD axial anomaly also plays an
important role in the spin structure of the proton.
In high-energy processes $B$ and $D_s$ meson decays 
to final states involving the $\eta'$ are enhanced
through coupling to gluonic intermediate states.
In low-energy reactions 
the $\eta' \rightarrow \eta \pi \pi$ decay process,
exclusive $\eta' \pi$ production in exotic odd-$L$ partial 
waves and $\eta'$ interactions with nuclear media 
are catalysed by OZI violation associated with the $\eta'$.
QCD inspired models 
are used to describe these different processes 
and the interplay of 
confinement, chiral symmetry and axial U(1) dynamics.
New data on possible $\eta$ and $\eta'$ bound states 
in nuclei
is expected soon from running and planned experiments at 
COSY, ELSA and GSI, and will help to further pin down the dynamics of axial U(1) symmetry breaking in low-energy QCD.

\vspace{0.5cm}

{\bf Acknowledgements} \\

I thank M. Faessler, V. Metag, P. Moskal and K. Suzuki 
for helpful communications. 
I thank M. Praszalowicz for the invitation to talk at 
this stimulating meeting. 
The research of S.D.B. is supported by the Austrian Science Fund, FWF, through grant P23753.
The research leading to these results has received funding from the European Union Seventh Framework Programme 
(FP7/2007-2013) under grant agreement number 283286.

\vspace{0.5cm}


\end{document}